\documentclass[pdflatex,sn-mathphys]{sn-jnl}

\jyear{2022}%
\usepackage{listings}
\usepackage[table,svgnames,dvipsnames]{xcolor}
\definecolor{codegreen}{rgb}{0,0.6,0}
\definecolor{codegray}{rgb}{0.5,0.5,0.5}
\definecolor{codepurple}{rgb}{0.58,0,0.82}
\definecolor{backcolour}{rgb}{0.95,0.95,0.92}
\lstdefinestyle{mystyle}{
    backgroundcolor=\color{backcolour},   
    commentstyle=\color{codegreen},
    keywordstyle=\color{orange},
    numberstyle=\tiny\color{codegray},
    stringstyle=\color{codepurple},
    basicstyle=\ttfamily\footnotesize,
    breakatwhitespace=false,         
    breaklines=true,                 
    captionpos=b,                    
    keepspaces=true,                 
    numbers=left,                    
    numbersep=5pt,                  
    showspaces=false,                
    showstringspaces=false,
    showtabs=false,                  
    tabsize=2
}
\lstdefinelanguage{json}{
    basicstyle=\normalfont\ttfamily,
    numbers=left,
    numberstyle=\scriptsize,
    stepnumber=1,
    numbersep=8pt,
    showstringspaces=false,
    breaklines=true,
    frame=lines
}
\lstdefinelanguage{JavaScript}{
  keywords={typeof, new, true, false, catch, let, function, return, null, catch, switch, var, if, in, while, do, else, case, break},
  keywordstyle=\color{blue}\bfseries,
  ndkeywords={class, export, boolean, throw, implements, import, this},
  ndkeywordstyle=\color{darkgray}\bfseries,
  identifierstyle=\color{black},
  sensitive=false,
  comment=[l]{//},
  morecomment=[s]{/*}{*/},
  commentstyle=\color{purple}\ttfamily,
  stringstyle=\color{red}\ttfamily,
  morestring=[b]',
  morestring=[b]"
}

\definecolor{delim}{RGB}{20,105,176}
\definecolor{numb}{RGB}{106, 109, 32}
\definecolor{string}{rgb}{0.64,0.08,0.08}

\lstdefinelanguage{JSON}{
    numbers=left,
    numberstyle=\tiny,
    frame=single,
    rulecolor=\color{black},
    showspaces=false,
    showtabs=false,
    breaklines=true,
    postbreak=\raisebox{0ex}[0ex][0ex]{\ensuremath{\color{gray}\hookrightarrow\space}},
    breakatwhitespace=true,
    basicstyle=\ttfamily\tiny,
    upquote=true,
    morestring=[b]",
    stringstyle=\color{string},
    literate=
     *{0}{{{\color{numb}0}}}{1}
      {1}{{{\color{numb}1}}}{1}
      {2}{{{\color{numb}2}}}{1}
      {3}{{{\color{numb}3}}}{1}
      {4}{{{\color{numb}4}}}{1}
      {5}{{{\color{numb}5}}}{1}
      {6}{{{\color{numb}6}}}{1}
      {7}{{{\color{numb}7}}}{1}
      {8}{{{\color{numb}8}}}{1}
      {9}{{{\color{numb}9}}}{1}
      {\{}{{{\color{delim}{\{}}}}{1}
      {\}}{{{\color{delim}{\}}}}}{1}
      {[}{{{\color{delim}{[}}}}{1}
      {]}{{{\color{delim}{]}}}}{1},
}
\theoremstyle{thmstyleone}%
\theoremstyle{thmstyletwo}%

\theoremstyle{thmstylethree}%

\begin{document}

\title[Efficient Chrome Extension for Phishing Detection Using ML Techniques]{Efficient Chrome Extension for Phishing Detection Using Machine Learning Techniques}


\author[1]{\fnm{Leand} \sur{Tha\c{c}i}}\email{leand.thaci@student.uni-pr.edu}

\author*[1]{\fnm{Arbnor} \sur{Halili}}\email{arbnor.halili@uni-pr.edu}

\author[2]{\fnm{Kamer} \sur{Vishi}}\email{kamerv@ifi.uio.no}
\author[1]{\fnm{Blerim} \sur{Rexha}}\email{blerim.rexha@uni-pr.edu}

\affil*[1]{\orgdiv{Faculty of Electrical and Computer Engineering}, \orgname{University of Prishtina}, \orgaddress{\street{Kodra e Diellit, p.n.}, \city{Prishtina}, \postcode{10000}, \country{Kosovo}}}

\affil[2]{\orgdiv{Department of Informatics}, \orgname{University of Oslo}, \orgaddress{\street{Gaustadalléen 23B}, \city{Oslo}, \postcode{N-0373}, \country{Norway}}}

                                           
\abstract{The growth of digitalization services via web browsers has simplified our daily routine of doing business. But at the same time, it has made the web browser very attractive for several cyber-attacks. Web phishing is a well-known cyberattack that is used by attackers camouflaging as trustworthy web servers to obtain sensitive user information such as credit card numbers, bank information, personal ID, social security number, and username and passwords. In recent years many techniques were developed to identify the authentic web pages that user visits and warn them when the webpage is a phish. In this paper, we have developed an extension for Chrome as the most favorite web browser, that will serve as a middleware between the user and phishing websites. The Chrome extension named “NoPhish” shall identify a phishing webpage based on several Machine Learning techniques. We have used the training dataset from “PhishTank” and extracted the 22 most popular features as rated by Alexa database. The training algorithms used are Random Forest, Support Vector Machine, and k-Nearest Neighbor. The performance results show that Random Forest delivers the best precision.}




\maketitle

\section{Introduction}
\label{sec:introduction}
Nowadays, cyber-attacks are considered to be the most prevalent attacks that threatens our security and privacy on the Internet. Starting from some simple actions up to critical cases like US Elections \cite{UsElectionAttacks}, cyber security has become a topic for every person and government around the world. 

According to the IBM Security X-Force Threat Intelligence Index report published in 2022, 41\% of all cyber-attacks came from phishing, making it the top 1 technique most used for initial attack vector \cite{IBM2022}. Comparing presence of this technique in report of 2021, there is a raise in 8 percent \cite{IBM2021}.  Phishing is a cyber-attack in which attackers try to steal information or sensitive data from victims, camouflaging as a reliable source. Additionally, phishing has also become a bridge for other attacks, tricking users into downloading malicious documents like viruses, malware, or ransomware.

\begin{figure}[ht]
\centerline{\includegraphics[width=0.7\columnwidth]{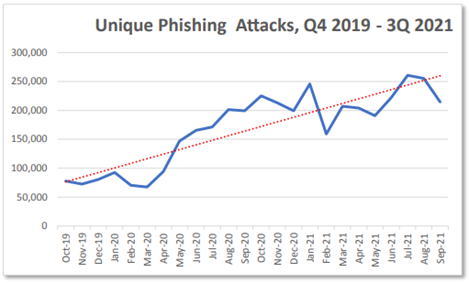}}
\caption{Statistics in a period of 2 years (growth is present all the time)~\cite{APWG}}
\label{APWG growth in phishing sites}
\end{figure}

\begin{figure}[ht]
\centerline{\includegraphics[width=0.6\columnwidth]{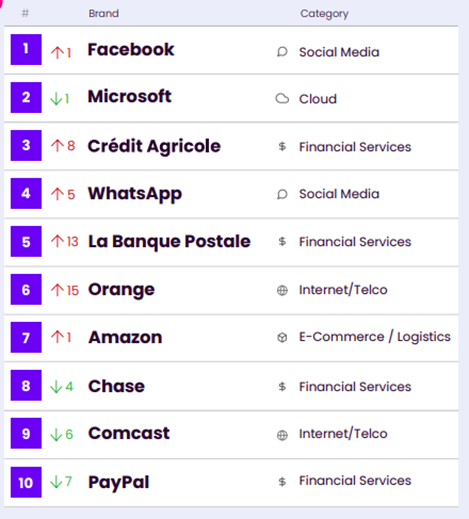}}
\caption{The ten most impersonated brands in phishing attacks~\cite{PhishersFavorites2021}}
\label{PshihersFavoritesTop10}
\end{figure}

One of our biggest motivations to further explore this topic was a gradual increase over the last two years of the number of phishing sites, as is presented in Figure \ref{APWG growth in phishing sites}. 
Victims of attacks can be from different categories, from large companies to individuals as ordinary Internet users. The most frequent attack on large companies is targeting the employees of that company through phishing emails. In terms of attacks which do not target a specific person or company, but are made for large masses, the most type of frequent attack is the imitation of large companies. According to statistics released by Vade Secure for the 2021 \cite{PhishersFavorites2021} and presented in Figure \ref{PshihersFavoritesTop10}, the company which is most often imitated is Facebook which overcame  "Microsoft", which was far away in first place. As of this writing, based on same report, Facebook has 2.8 billion users, a large pool of potential victims ready to bite on Facebook phishing messages and web pages. There is also a large number of phishing attacks that don’t impersonate Facebook directly. Instead, the emails impersonate another brand and include links to Facebook phishing pages, exploiting user trust in the Facebook brand.

\begin{figure}[ht]
\centerline{\includegraphics[width=0.7\columnwidth]{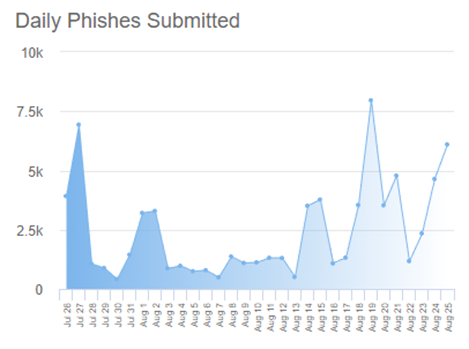}}
\caption{Statistics of daily reports for phishing sites in PhishTank~\cite{PhishTank}}
\label{PhishTank daily reports}
\end{figure}

The battle between hackers and defenders on this topic is on daily basis, and the result is never clear. There is no clear index or methodology specified that would classify who is dominating the Internet ecosystem however we are witnesses of such a situation.  This battle is best illustrated in Figure \ref{PhishTank daily reports} where statistics of daily reporting within August 2022 for phishing are presented.

Based on this growth of phishing sites we evaluate that there is always room for improvement in terms of defense against these attacks and any research in this regard means fewer victims. Obviously, attacks of this kind cannot be solved once and for all, constant efforts must be made to study new methods of attack and to adapt protection against them. In this paper, we tend that through detailed research and analysis to provide new and efficient solutions for application by using the latest technologies and techniques. In addition to the tool development and training part, this paper gives importance also to the simplicity of using the tool. The goal is that every user regardless of their professional background in this field, be able to get the right information about the site being visited.

Work done for this solution combines knowledge from different fields of study, like Machine Learning, Web programming, and Web security. Machine Learning (ML) is seen as a mechanism that will enrich our tool with a way of learning and deciding based on a model that is created. The model will be created based on training data. Similar usage was seen earlier in a wide variety of applications, such as email filtering and computer vision, where it is difficult or impossible for conventional algorithms to perform the necessary tasks. 

This paper's purpose is to propose a tool that can classify phishing sites. Classification falls under the Supervised Learning paradigm, where the model is trained with labeled data, a process that is often quite costly in all respects. Algorithms that will be discussed and that belong to this paradigm are:
\begin{itemize}
\itemsep0em
    \item \textbf{SVM} (\textbf{Support Vector Machine}) - As it is presented in Figure \ref{SVM} the classification boundaries are defined by the so-called support vectors. This is the perfect case when the data is sufficiently differentiable and there are no exceptions (outliers), respectively instances which are located outside the defined boundaries. If the model is trained using hard margins (hard margin classification) then the model works only if the data are linearly separated. For this reason, more flexible parameters of margins should be defined, respectively soft margin classification.
    \begin{figure}[ht]
\centerline{\includegraphics[width=1\columnwidth]{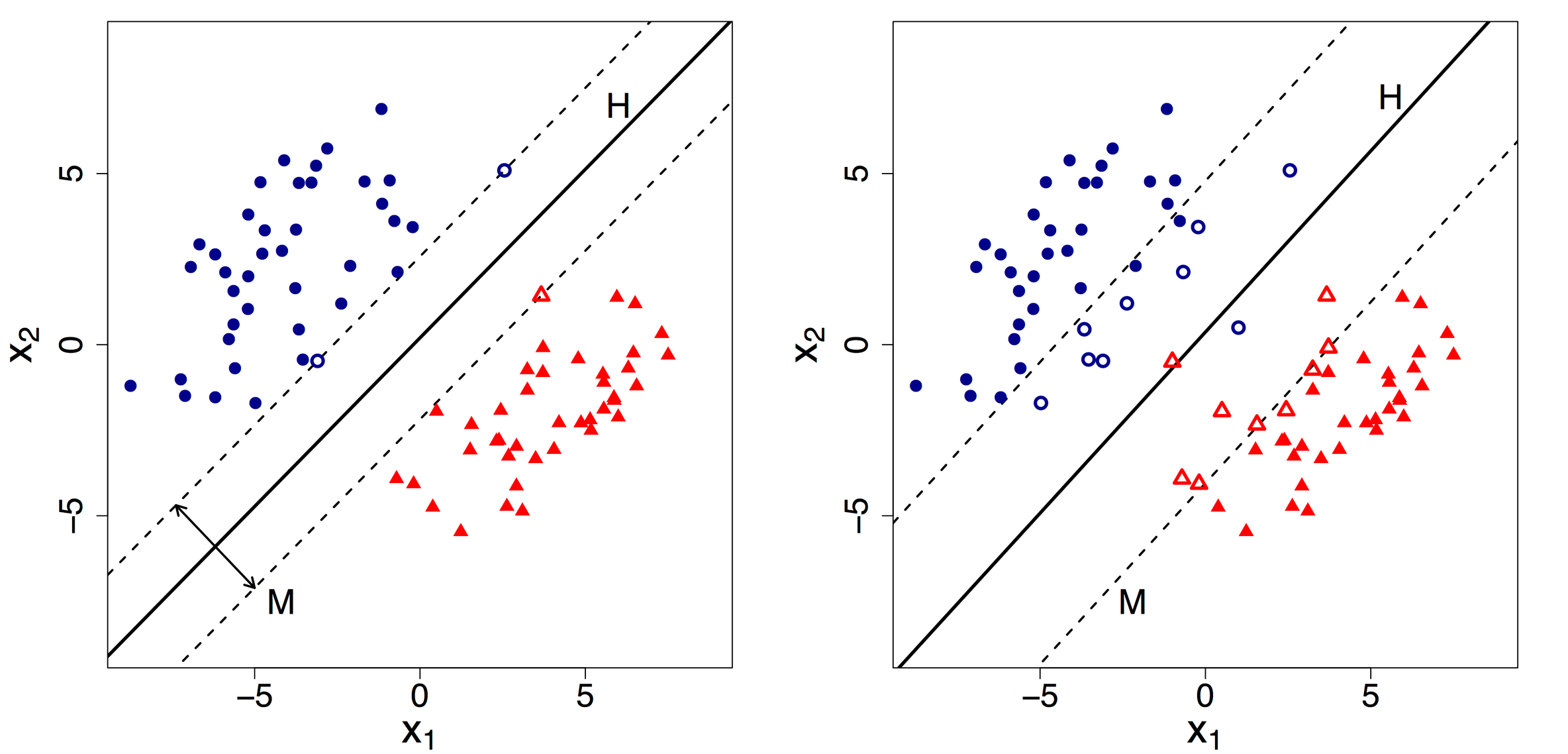}}
\caption{Differences between soft and hard margins for SVM ~\cite{SVM}}
\label{SVM}
\end{figure}

    \item \textbf{kNN (k-Nearest Neighbors)} - The principle behind kNN is to find a predetermined number of training samples closest to the distance from the new point and to predict from these. The number of samples can be a user-defined constant (k-nearest neighbor learning), or variable based on local point density (radius-based neighbor learning). Distance, in general, can be any metric mass: the standard Euclidean distance is the most common choice. Neighbor-based methods are known as non-generalized Machine Learning methods, as they simply "recall" all their training records.
    
    Despite its simplicity, kNN has been successful in a large number of classification and regression problems, including handwritten figures and satellite image scenes. Being a non-parametric method, it is often successful in classification situations where the decision limit is very irregular.
    
    \item \textbf{Random Forest} - Is a flexible, easy-to-use algorithm that produces, even without adjusting parameters for specific cases, an excellent result most of the time. Random Forest is one of the most widely used algorithms, due to its simplicity and diversity (can be used for both classification and regression tasks). Random Forest is an advanced version of the Decision Trees algorithm, it basically uses the Decision Trees algorithm in different subgroups of the dataset and uses the average value to increase performance and avoid over fitting. This method is known as "bagging" or "bootstrap aggregating". Suppose we fit a model to our training data Z = {(x1, y1), (x2,y2),.., (xN, yN)}, obtaining the prediction $\hat{f}$(x) at input x. Bootstrap aggregation or bagging averages this prediction over a collection of bootstrap samples, thereby reducing its variance. For each bootstrap sample \(Z^{*b}b\), b = 1, 2, ... ,B, we fit our model, giving prediction $\hat{f}*^{b}(x)$~\cite{hastie2009elements}. The bagging estimate is defined by

\begin{equation}
     \hat{f}_{bag}(x) = \frac{1}{B} \sum_{b=1}^{B} \hat{f}^{*b}(x)
\end{equation}

Figure \ref{RandomForestDiagram} presents the complete diagram of the Random Forest algorithm.

\begin{figure}[ht]
\centerline{\includegraphics[width=0.8\columnwidth]{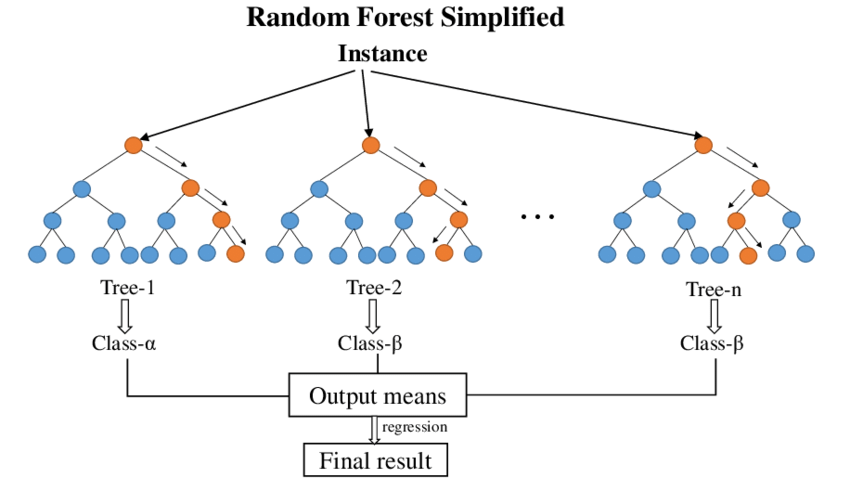}}
\caption{Random Forest Diagram~\cite{RandomForestDiagram}}
\label{RandomForestDiagram}
\end{figure}
\end{itemize}

One of the first challenges when implementing Machine Learning in identifying phishing sites is the possession of a dataset which is updated with the latest data and has real data. Fortunately, there are many online databases with multiple datasets. One of the biggest anti-phishing company, where thousands of pages are submitted and validated every day is "PhishTank"~\cite{PhishTank}. Among useful information and statistics, this web page also has an API through which developers can obtain data.

The rest of the paper is organized as follows. The next section presents background information on phishing attacks. Section \ref{sec:NoPhish} describes the proposed model, namely NoPhish, that is developed to detect malicious websites. Section \ref{sec:experimental_results} details the evaluation results of the model. Finally, Section \ref{sec:conclusion} concludes the paper.

\section{Related Work}
\label{sec:relatedwork}
Sadly, in the battle against phishing attacks, the attacking side is always one step ahead, so the defensive side needs always to keep up with the new methods of attacking \cite{alkhalil2021phishing}. The process of constant sophisticating the tools and techniques for protecting users from phishing attacks is forcing attackers to adapt and evolve their techniques. Today phishing attacks are becoming more and more polymorphic \cite{jcp1040037}. The situation becomes even worse when it is known that the most perfect attacks occur within minutes and do not recur twice by the attacker without improving for the next time. The ''zero-day attacks'' that by definition are attacks that have not happened before, today are not uncommon \cite{Elie2019}. These attacks are always launched with a new set of IP addresses and a new variant of the virus. 

The most common method of detecting phishing sites is by updating URLs and IPs addresses listed as phishing in the antivirus database, this method is known as ''Blacklist'' \cite{aljofey2022effective}. To avoid blacklists attackers use creative techniques to cheat the system by modifying the URL to look legitimate through obfuscation and many techniques other simple methods including the ''fast-flux'' method \cite{9484614}. Through this method proxies 
are automatically generated to host the web page and new URLs are generated through algorithms. The main drawback of this method is that, it can not detect ''Zero-day attacks''. Thus, there must be at least one victim who will report this site and will blacklist that page. One of the most popular tools of this type is ''Netcraft Extension'' \cite{NetCraftExtension}.
This gadget consists of a large community, and expects the first victim to report a page in order to inform other users not to visit that page. 

Another method for detecting phishing sites is a system with predefined rules. Where the tool analyzes the content of a site and based on the rules it has predefined and decides whether that site is considered phishing or not \cite{lin2021phishpedia}. The problem with this type of system is, although in principle it can stop "zero-day" attacks, predefined features are not guaranteed to be found in every attack so the number of false positives is very high and usage remains very narrow. The most popular tool of this type of system is "Phish Detector"\cite{PhishDetector}. A tool that prevents phishing attacks in online banking. It is a "rule-based" system and claims zero false negative predictions. The disadvantage is that it is limited to the domain of online banking. 

The third method, detecting phishing sites using Machine Learning, tends to stop "zero-day" attacks, including as many attacking domains as possible. 
Nowadays there are plenty of tools with this method, but there is always room for improvement. One of the early tools of this kind is the "Cascaded Phish Detector" which similar to this paper
consists of a client-side component and a server-side component \cite{CascadedPhishDetector}. This gadget  only takes into account the HTML content of the page, we think there are also other kind of features that can be extracted from a webpage and are informative as mentioned in Section \ref{sec:introduction}.


The authors in \cite{9214225} used Machine Learning to predict phishing. We find missing in this paper that there is no explanation of the ratio used to divide the training and testing set. Furthermore, there is no concrete implementation of this algorithm; it is not part of any tool, or we have not found any concrete implementation and ease of usage, which is one of the critical points of our work.

An interesting approach is found at \cite{Gowtham2014}. Here an architecture which is mainly referred to a preprocessing of data is proposed. However, it only took into consideration one algorithm and could lead to a over fitting.

\cite{Aphishdetectorusinglightweightsearchfeatures} used a technique of search for classification of a web page as phishing or not. Pseudocode presented on the paper shows that algorithm instead of learning is using services through API and direct requests to make a decision. In this case Google search engine has the main role since top six Google search results are used to make decision. It worth mentioning that in most cases, top six Google search results change very rare, hence, it could present a high potential for wrong 
classification.

We found interesting the work of \cite{Aphishdetectorusinglightweightsearchfeatures}, which ranked Random Forest to be best (together with XGboost) among 12 algorithms used. However, even here there is no concrete implementation to link this result with a concrete tool.

Furthermore, from our research made on current work and state of the art, best to our knowledge, we could identify a research gap in this intensive field. There are three principles that we took into consideration:
\begin{itemize}
\itemsep-0,3em
    \item effectiveness of the solution,
    \item easiness of implementation and usage, and
    \item innovation.
\end{itemize}
From studies and work done until now we found solutions that are effective (like \cite{Gowtham2014} but there is no implementation or concrete use. We found work that was easy to use \cite{9214225} but not enough effective. Our solution tends to fulfill all these three principles and tends to do it in a novel, innovative way, nor presented until now by any author.

\section{Our Approach - NoPhish}
\label{sec:NoPhish}
Since the classification of whether the site is suspicious or not will be done by the tool trained with Machine Learning, the question arises on what the tool will base its prediction. The main thing to look for when detecting suspicious phishing sites is the URL of the site. Nowadays it is not enough just to look at whether the site uses the secure HTTPS protocol for communication. According to the latest report from APWG (Anti-Phishing Working Group), about 80\% of phishing sites use the protocol HTTPS \cite{APWG}. Therefore, some properties are extracted from the URL based on which a page can be considered suspicious. Other features are derived from the content of the site, where the authors of these sites often focus only on copy the look of the real page and leave many flaws which can be identified.
And the last category of site features are extracted from  searching this site in the most popular databases of statistics about web pages, such as the Alexa Database \cite{AlexaInternet}. 
The most important part of the tool is definitely the training algorithm, based on which the tool
makes decisions about how to classify the pages it handles. Within Machine Learning we have a large diversity of algorithms which are adequate for different situations. For this paper the variety is
narrowed down to the 3 most adequate algorithms, which will be tested and compared. They are: Random
Forest, SVM (Support Vector Machine), kNN (k-Nearest Neighbor). Chrome Extension will play the role of the middle man between the user and the classifier.
Google Chrome offers the ability to add a tool
add-on which can be used on any page you browse, it is called a Chrome Extension. The question is, why exactly Google Chrome? According to
November 2020 statistics ~\cite{W3Counter}, Chrome browser dominates with 65.3\% of worldwide use.

The tool \footnote{NoPhish Repository: \href{https://github.com/LeandThaqi/NoPhish}{https://github.com/LeandThaqi/NoPhish}} consists of two main components: the "client-side" component with which the user will interact and, the "server-side" component which will be the classifier that decides whether a site is considered phishing and the web server which will play the role of mediator between the two components and will create the environment for the execution of the classifier.

\begin{figure}[ht]
\centerline{\includegraphics[width=1\columnwidth]{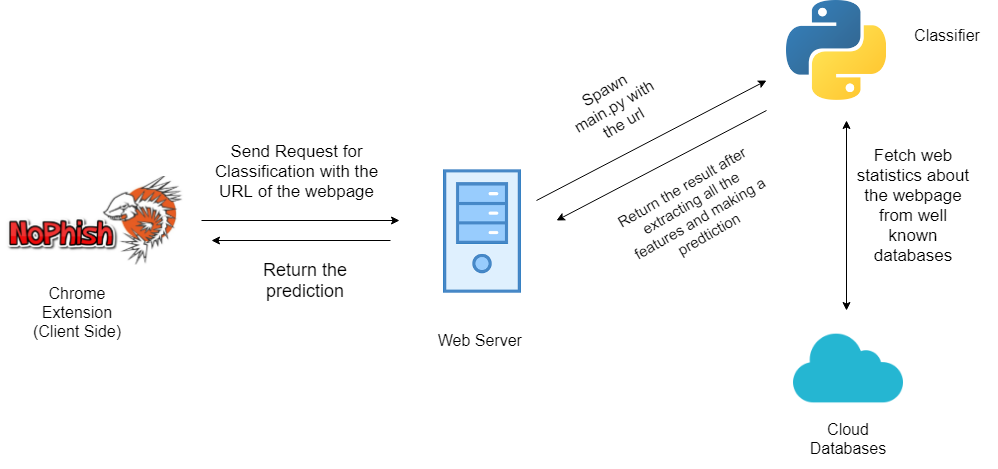}}
\caption{NoPhish architecture}
\label{ToolArchitecture}
\end{figure}
Since the key point of this paper is Machine Learning, it was decided to use the Python language (version 3.9) for the implementation of the classifier component. One of the advantages of this language is definitely the large number of libraries. The Python language is today recognized as a leader in the field of Machine Learning, through the skicit-learn library the implementation of algorithms for Machine Learning is easier than ever. In addition Python also simplifies the data manipulation procedure for extraction of site features. The numpy library is used for manipulations with large multidimensional arrays, while matplotlib is used for graphical representations of results and statistical data needed for visualization. The "client-side" component will be developed as "Chrome Extension" for very easy user access. The reason for selecting the Google Chrome browser is the large percentage of users in the market. The Webserver who will make the link between the two
components will be developed in the Node.Js language, the reasons being good performance and the ease of use.

Today the Internet world is rich in data. Many different sites contribute to the fight against phishing with different datasets. The dataset selected for the tool training for this paper was donated to the University of California, Irvine \cite{Dataset}. It was originally donated with data for about 2400 pages in 2015 but has been constantly updated and today this number has reached up to about 11000. The database contains data on thirty page features.

\subsection{Determining the features based on which pages will be classified}
Because the dataset contains features that help detect sites not only for phishing but also for other risks. Twenty-two of the thirty features listed bellow are considered to be phishing information.

\begin{enumerate}[start=0]
\itemsep0em
  \item \textbf{Using IP address instead of domain} - If an IP address is used as an alternative of the domain name in the URL, such as “http://125.98.3.123/fake.html”, users can be sure that someone is trying to steal their personal information. Sometimes, the IP address is even transformed into hexadecimal code as shown in the following link “http://0x58.0xCC.0xCA.0x62/2/paypal.ca/index .html”. 
  
  \item \textbf{Long URL to Hide the Suspicious Part} - Phishers can use long URL to hide the doubtful part in the address bar

  \item \textbf{Using URL Shortening Services “TinyURL”} - URL shortening is a method on the “World Wide Web” in which a URL may be made considerably smaller in length and still lead to the required webpage. This is accomplished by means of an “HTTP Redirect” on a domain name that is short, which links to the webpage that has a long URL.
  \item \textbf{URL’s having “@” Symbol} - Using “@” symbol in the URL leads the browser to ignore everything preceding the “@” symbol and the real address often follows the “@” symbol
  \item \textbf{Redirecting using “//”} - The existence of “//” within the URL path means that the user will be redirected to another website. 
  
  \item \textbf{Adding Prefix or Suffix Separated by (-) to the Domain} - The dash symbol is rarely used in legitimate URLs. Phishers tend to add prefixes or suffixes separated by (-) to the domain name so that users feel that they are dealing with a legitimate webpage
  \item \textbf{Sub Domain and Multi Sub Domains} - If a page has many subdomains then it can be considered that the attacker is trying to disguise the site domain.
  \item \textbf{Domain Registration Length} - Based on the fact that a phishing website lives for a short period of time, we believe that trustworthy domains are regularly paid for several years in advance. 
  \item \textbf{Favicon} - A favicon is a graphic image (icon) associated with a specific webpage. Many existing user agents such as graphical browsers and newsreaders show favicon as a visual reminder of the website identity in the address bar. If the favicon is loaded from a domain other than that shown in the address bar, then the webpage is likely to be considered a Phishing attempt. 
  \item \textbf{The Existence of “HTTPS” Token in the Domain} - The phishers may add the “HTTPS” token to the domain part of a URL in order to trick users. 
  \item \textbf{Request URL} - Request URL examines whether the external objects contained within a webpage such as images, videos and sounds are loaded from another domain. In legitimate webpages, the webpage address and most of objects embedded within the webpage are sharing the same domain. 
  \item \textbf{URL of Anchor} - An anchor is an element defined by the <a> tag. This feature is treated exactly as “Request URL”. However, for this feature we examine: 
\begin{enumerate}
    \item If the <a> tags and the website have different domain names,
    \item If the anchor does not link to any webpage.
\end{enumerate}
  \item \textbf{Links in <Meta>, <Script> and <Link> tags} - Given that our investigation covers all angles likely to be used in the webpage source code, we find that it is common for legitimate websites to use <Meta> tags to offer metadata about the HTML document; <Script> tags to create a client side script; and <Link> tags to retrieve other web resources. It is expected that these tags are linked to the same domain of the webpage. 
  \item \textbf{Server Form Handler (SFH)} - SFHs that contain an empty string or “about:blank” are considered doubtful because an action should be taken upon the submitted information. In addition, if the domain name in SFHs is different from the domain name of the webpage, this reveals that the webpage is suspicious because the submitted information is rarely handled by external domains. 
  \item \textbf{Submitting Information to Email} - Web form allows a user to submit his personal information that is directed to a server for processing. A phisher might redirect the user’s information to his personal email. To that end, a server-side script language might be used such as “mail()” function in PHP. One more client-side function that might be used for this purpose is the “mailto:” function
  \item \textbf{Abnormal URL} - This feature can be extracted from WHOIS database. For a legitimate website, identity is typically part of its URL. 
  \item \textbf{IFrame Redirection} - IFrame is an HTML tag used to display an additional webpage into one that is currently shown. Phishers can make use of the “iframe” tag and make it invisible i.e. without frame borders. In this regard, phishers make use of the “frameBorder” attribute which causes the browser to render a visual delineation. 
  \item \textbf{Age of Domain} - This feature can be extracted from WHOIS database (Whois 2005). Most phishing websites live for a short period of time. 
  \item \textbf{DNS Record} - For phishing websites, either the claimed identity is not recognized by the WHOIS database (Whois 2005) or no records founded for the hostname (Pan and Ding 2006). If the DNS record is empty or not found then the website is classified as “Phishing”, otherwise it is classified as “Legitimate”. 
  \item \textbf{Website Traffic} - This feature measures the popularity of the website by determining the number of visitors and the number of pages they visit. 
  \item \textbf{Google Index} - This feature examines whether a website is in Google’s index or not. When a site is indexed by Google, it is displayed on search results (Webmaster resources, 2014). 
  \item \textbf{Statistical-Reports Based Feature} - Various companies like PhishTank or StopBadware form statistical reports exposing the IPs and domains most used for phishing. This feature checks if the host is part of these reports.
\end{enumerate}
\subsection{Choosing the most efficient algorithm for our specific case}
\begin{figure}[ht]
\centerline{\includegraphics[width=0.85\columnwidth]{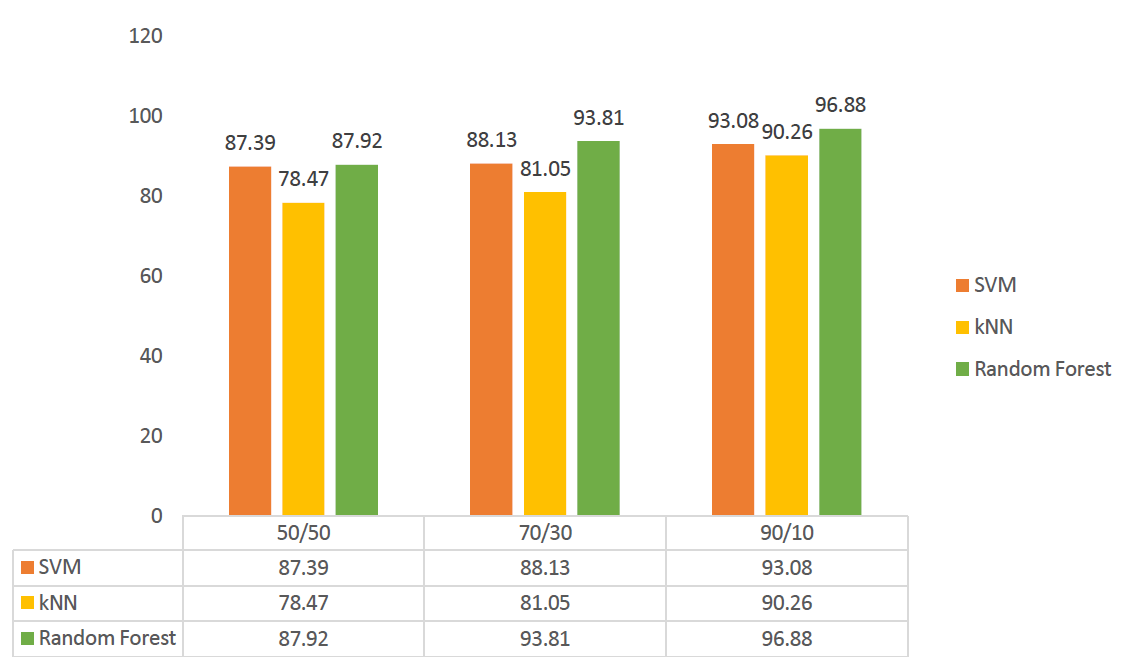}}
\caption{Performance results for accuracy from the three algorithms}
\label{PerformanceTesting}
\end{figure}

For this paper we took under consideration three classification algorithms, they are: SVM, kNN and Random Forest. All will be tested and the algorithm that performs the best will be selected. Figure \ref{PerformanceTesting} shows the results of the accuracy of the algorithms from tests by dividing the dataset into different proportions (training/testing): 50/50, 70/30, 90/10.

Based on the extracted statistics it can be seen that the Random Forest algorithm outperforms other algorithms in this experiment in terms of accuracy. The efficiency of algorithms in ML is also measured by several other parameters such as: precision score, recall score and confusion matrix.


\begin{table}[tbp]
\centering
\caption{Performance results for precision, recall, and confusion matrix\\}
\begin{tabular}{p{1cm}clllll} 
 & \textbf{Precision} & \textbf{Recall} & \textbf{TN}  & \textbf{FP} & \textbf{FN} & \textbf{TP}   \\ 
\hline\hline
\rule{0pt}{3ex} \textbf{SVM}   & 0.913     & 0.969  & 422 & 57 & 19 & 601  \\ 
\hline
\rule{0pt}{3ex}  \textbf{kNN}  & 0.901     & 0.904  & 431 & 48 & 59 & 561  \\ 
\hline
\noalign{\vskip 1.5mm} \textbf{Random Forest} & 0.921 & 0.970  & 435 & 44 & 17 & 603
\end{tabular}
\end{table}

Based on the test results it is seen that Random Forest is the most accurate algorithm for this case. And considering the "bagging" feature of this algorithm which mitigates the problem of overfitting, the tool will be developed with the Random Forest algorithm. The following code snippet trains the model with the Random Forest Algorithm, this model will be used to predict if the websites are phishing (Listing \ref{lst:modelTraining}).

\begin{lstlisting}[language=Python, caption=The code that trains the model using the Random Forest Algorithm, label={lst:modelTraining}]
clf4 = RandomForestClassifier(verbose=2, oob_score=True)
clf4.fit(features_train, labels_train)
predictions = clf4.predict(features_test)
accuracy = 100.0 * accuracy_score(labels_test, predictions)
importances = clf4.feature_importances_
std=np.std([tree.feature_importances_ for tree in clf4.estimators_],axis=0)
indices = np.argsort(importances)[::-1]


\end{lstlisting}

\subsection{Feature Importance}
\begin{figure}[ht]
\centerline{\includegraphics[width=1\columnwidth]{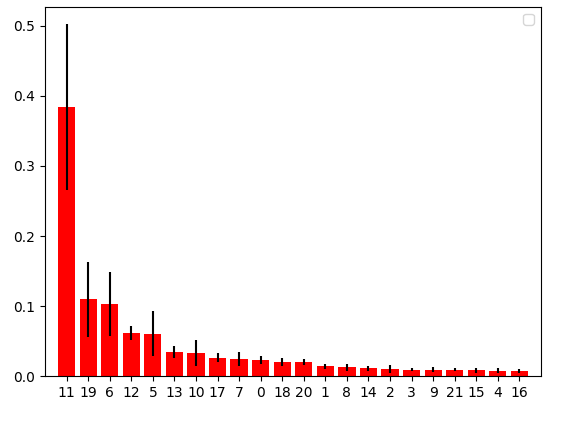}}
\caption{Importance of features based on MDI}
\label{MDIFeatureImportance}
\end{figure}

The Random Forest algorithm also offers us the importance of each feature from the training. Through the \textit{feature\_importance\_object}, the features are ranked based on how informative they were during the training. These percentages are calculated by the algorithm from MDI (Mean Decrease in Impurity) so as that feature reduces the impurity of the dataset. Figure \ref{MDIFeatureImportance} presents the ranking of the features according to the MDI.

\begin{figure}[ht]
\centerline{\includegraphics[width=0.9\columnwidth]{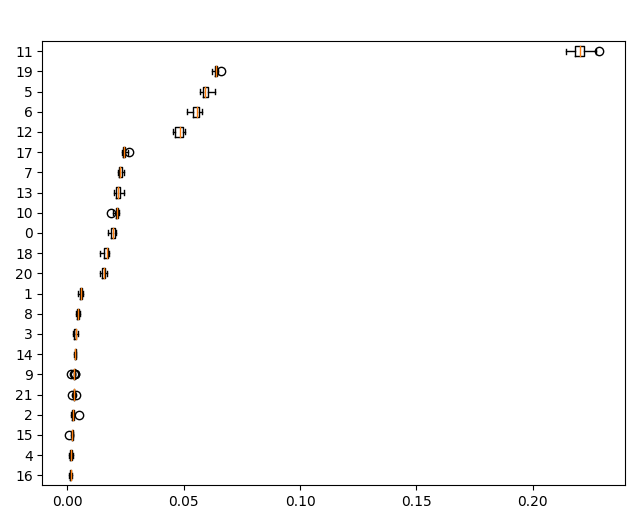}}
\caption{Importance of features based on permutation}
\label{PermutationBasedImportances}
\end{figure}

A problem that has emerged recently is that the importance of features based on impurity can inflate the importance of numerical features. Furthermore, the impurity-based feature importance of random forests suffers from being computed on statistics derived from the training dataset: the importances can be high even for features that are not predictive of the target variable, as long as the model has the capacity to use them to overfit. Alternatively, the importance of features can be calculated from permutation importances as illustrated in Figure \ref{PermutationBasedImportances}. 

As we can see from both the graph extracted from different methods that \textit{feature 11} which is the URL of the anchor tags is the most important feature. Logically the danger of phishing sites is the redirection into suspicious sites.

\subsection{Client-side Component (The face of the tool)}
The component with which the user will interact will be developed as a Chrome Extension. Chrome Extension development is simple, the three elements needed for a functional extension are: 
\begin{enumerate}
\itemsep-0,3em
    \item \textit{popup.html} - where HTML elements are defined,
    \item \textit{popup.js} - where functions are written in javascript, and
    \item  \textit{manifest.json} - where extensions details are written. The following code snippet represents the manifest.json file (Listing \ref{lst:manifest}).
\end{enumerate}

\begin{lstlisting}[language=JSON, caption=manifest.json (chrome extension configuration file), label={lst:manifest}]
{
    "manifest_version": 2,
  
    "name": "NoPhish",
    "description": "This extension will analyze a page and predict if it is a phishing site or not , by using machine learning!",
    "version": "0.8",
    "browser_action": {
     "default_icon": "icon.png",
     "default_popup": "popup.html"
    },
    "icons": { "16": "icon.png",
        "48": "icon48.png",
       "128": "icon128.png" },
    "permissions": [
     "activeTab"
     ]
  }
\end{lstlisting}
The main purpose of Chrome Extension is to be as simple as possible.

\begin{figure}[ht]
\centerline{\includegraphics[width=1\columnwidth]{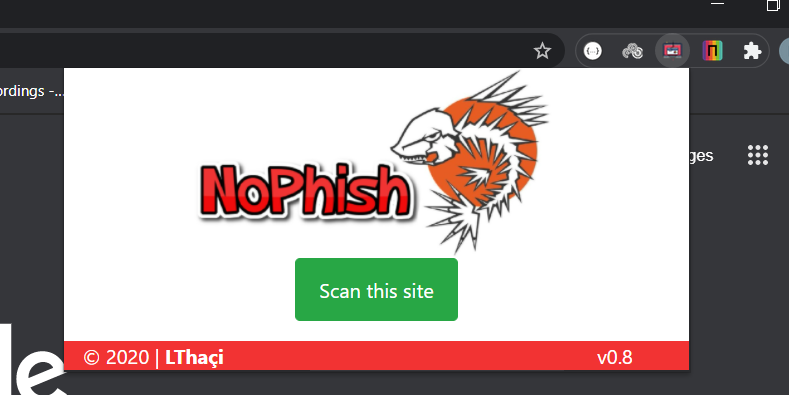}}
\caption{First view of the tool}
\label{FirstViewOfTheTool}
\end{figure}
\begin{figure}[ht]
\centerline{\includegraphics[width=1\columnwidth]{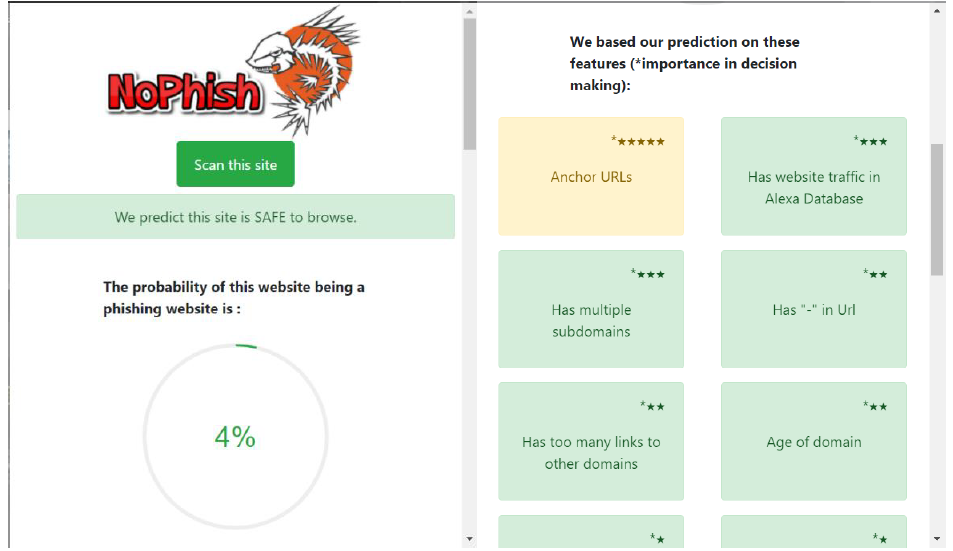}}
\caption{The safe prediction information displayed}
\label{ToolViewAfterPrediction}
\end{figure}

The design is made in a manner that with a click of a button all the necessary information is obtained. Figure \ref{FirstViewOfTheTool} shows the first view of the tool.
The "Scan this site" button sends a request to the web server with the URL of the site, where user is currently located and waits for a response. The following code snippet sends the request to the web server (Listing \ref{lst:sendingRequest}). 
\begin{lstlisting}[language=javascript, caption=Sending the request to check if the website which the user is browsing is phishing, label={lst:sendingRequest}]
      chrome.tabs.getSelected(null, function(tab) {
            let xhr = new XMLHttpRequest();
            let url = "http://localhost:3000/detectphishing";
            xhr.open("POST", url, true);
            xhr.setRequestHeader("Content-Type", "application/json");
            var data = JSON.stringify({"url": tab.url});
            xhr.send(data);
            ..
      });

\end{lstlisting}
The webserver accepts the request (This can be seen in the following code snippet \ref{lst:test}) , spawns the script which predicts if the site is phishing (This can be seen in the following code snippet \ref{lst:predictionScript}) then returns the response , as presented in Figure \ref{ToolViewAfterPrediction}, the prediction is initially displayed which notifies the user whether the tool evaluates the visited site as phishing or not. The calculated probability of how suspicious the site is is displayed and also for each feature of the site it is indicated which one failed during testing.

\begin{lstlisting}[language=javascript, caption=Webserver accepting a request and spawning the python script which evaluates if the website is phishing, label={lst:test}]
         get(url).then((pageContent) => {
          fs.writeFileSync('../innerHTML.txt',pageContent)
          const pythonProcess = spawn('python',["../main.py",url]);
          pythonProcess.stdout.on('data', (data) => {
            response.writeHead(200, {'Content-Type': 'application/json'});
            response.end(data.toString('utf8'));
          });
\end{lstlisting}

\begin{figure}[ht]
\centerline{\includegraphics[width=1\columnwidth]{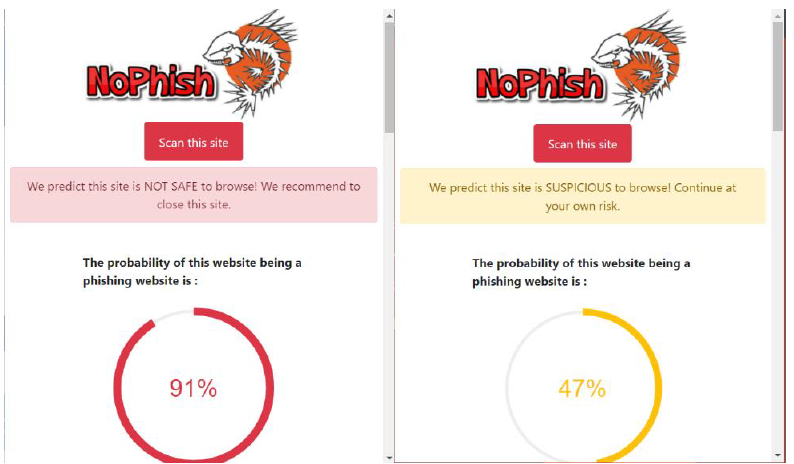}}
\caption{The dangerous and warning prediction information displayed}
\label{ToolViewAfterPredictionDangerAndWarning}
\end{figure}

\begin{lstlisting}[language=Python, caption=The python script which predicts if the website if phishing, label={lst:predictionScript}]
  def main():
    url = sys.argv[1]
    features_test = features_extraction.main(url)
    # 2d array per scikit-learn
    features_test = np.array(features_test).reshape((1, -1))
    clf = joblib.load(LOCALHOST_PATH + DIRECTORY_NAME + '/classifier/random_forest_new_4.pkl')
    prediction = clf.predict(features_test)


\end{lstlisting}
Based on the principle that a false positive prediction is less harmful than a false negative prediction, in addition to predicting whether it is phishing or not, a warning area has been added. This area indicates that the site is predicted to be secure but there are questionable signs in some features, so attention is required from the user to review which site it is on. Besides when the prediction returns safe as presented in Figure \ref{ToolViewAfterPrediction}, two other cases are presented in Figure \ref{ToolViewAfterPredictionDangerAndWarning}.

\section{Experimental Results}
\label{sec:experimental_results}
In order to test our tool against current solutions to this problem we did some experimental work. First we selected our testing subject which will be twenty-seven websites, fourteen of them are verified phishing websites taken from PhishTank\cite{PhishTank} and the other thirteen are well known websites proven that they dont have any malicious intent. 

\begin{figure}[ht]
\centerline{\includegraphics[width=1\columnwidth]{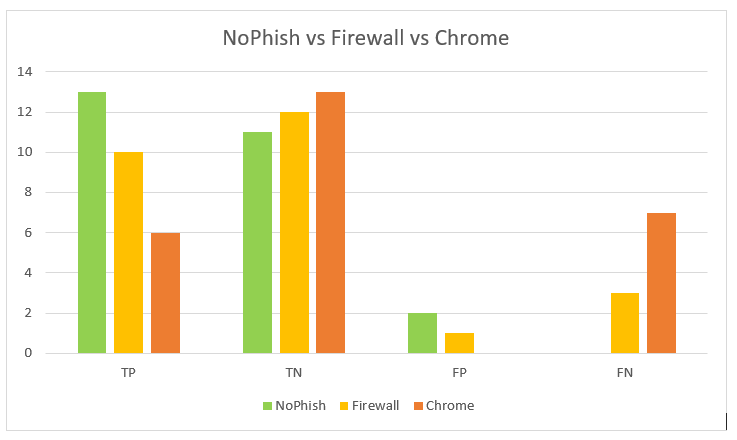}}
\caption{Statistics based on the confusion matrix}
\label{NoPhishVSFirewall}
\end{figure}

The tools that we did compare our work with were: Google Chrome`s built in firewall and the firewall of a world wide agency whose name will be undisclosed for privacy reasons. The methodology of testing was as follows, each website was scanned with the three tools. In the end we gathered all the information and displayed it as the confusion matrix(as presented in Figure \ref{NoPhishVSFirewall}) which we think is the best indicator of accuracy for this particular case.

Analyzing the results presented in Figure \ref{NoPhishVSFirewall}, one can conclude that NoPhish is the most efficient tool out of these three. If one analyzes the TP (True Positive) cases, one can conclude that NoPhish gives the best result, which means it detected more phishing sites than the other two tools. However, when one analyzes about TN (True Negative) and FP (False Positive) cases, it is easy to conclude that NoPhish is performing a bit worse comparing with other two mechanisms. NoPhish was found to have the least number of TN cases and the highest number of FP cases (even though it is a low number, difference is only one). This is a conservative part of our tool,  always trying to put the client on the safe side. The reason for this behaviour is that NoPhish besides the safe and dangerous prediction has a warning zone. In warning zone are all links, that NoPhish technically predicts as safe but are very near to the threshold, thus NoPhish warns the user. These warning predictions were counted as TP. This method allows NoPhish to have zero FN (False Negative) predictions and rates NoPhish above the other two tools taken into comparison.

\section{Conclusion}
\label{sec:conclusion}

From the evaluation results, one can conclude that developing a classifier by applying Machine Learning gives satisfactory results. Besides good results, there is a great potential for more extensive research in machine learning and phishing detection techniques. 

This paper's results are satisfactory by achieving high accuracy with a small number of false-negatives. Furthermore, the Random Forest algorithm proved more efficient in this research; it does not mean that there could not be another more successful algorithm in the future in other circumstances with extensive evolution of attacks. The features on which the forecast is based should always represent the state-of-the-art attack properties. Such as the number four feature where if the URL contains "//" it is considered suspicious, and it can be removed in the future. Because some browsers nowadays have already stopped redirecting to other pages if the URL contains "//", but as long as this security "bypass" is present in all browsers can not be neglected.

\section{Declarations}
\label{sec:declarations}

\textbf{Ethical Approval} \newline
Not applicable. \newline\newline
\textbf{Competing interests} \newline
Personal nature interest (continuous work on security research)\newline\newline
\textbf{Authors' contributions} \newline
A.H. wrote the introduction sections text, contributed to related work, and the main manuscript text.
K.V. wrote related work and contributed to the main manuscript text.
L.TH. wrote mostly on the main manuscript and also dealt with experimental results.
B.R. initiated the idea by writing the abstract and conclusion and was the main reviewer.
All authors reviewed the manuscript.\newline\newline
\textbf{Funding} \newline
Not applicable. \newline\newline
\textbf{Availability of data and materials} \newline
https://github.com/LeandThaqi/NoPhish

\bibliography{sn-article}


\end{document}